\RequirePackage{fix-cm}
\documentclass[twocolumn]{article}       
\usepackage{graphicx}
\usepackage{subfigure}
\usepackage{amsmath}
\usepackage{xspace}
\usepackage{algorithm}
\usepackage{algorithmic}
\usepackage{multirow}
\usepackage{authblk}

\newcommand{\algorithmicinput}{\textbf{Input:}}
\newcommand{\algorithmicoutput}{\textbf{Output:}}

\newcommand{\algorithmicdescription}{\textbf{Description:}}

\newcommand{\INPUT}{\item[{\algorithmicinput}]$\phantom{1}$\\}
\newcommand{\OUTPUT}{\item[\algorithmicoutput]$\phantom{1}$\\}

\newcommand{\DESCRIPTION}{\item[\algorithmicdescription]$\phantom{1}$\\}

\newtheorem{myDef}{Definition}
\begin{document}

\title{\textbf{ColluEagle: Collusive review spammer detection using Markov random fields}
}


\author[1]{Zhuo Wang \thanks{Corresponding author: zhuowang@sylu.edu.cn}}
\author[1]{Runlong Hu}
\author[1]{Qian Chen}
\author[1]{Pei Gao}
\author[2]{Xiaowei Xu}
\affil[1]{School of Information Science and Engineering, Shenyang Ligong University}
\affil[2]{University of Arkansas at Little Rock}


\maketitle

\begin{abstract}
Product reviews are extremely valuable for online shoppers in providing purchase decisions. Driven by immense profit incentives, fraudsters deliberately fabricate untruthful reviews to distort the reputation of online products. As online reviews become more and more important, group spamming, i.e., a team of fraudsters working collaboratively to attack a set of target products, becomes a new fashion. Previous works use review network effects, i.e. the relationships among reviewers, reviews, and products, to detect fake reviews or review spammers, but ignore time effects, which are critical in characterizing group spamming.
In this paper, we propose a novel Markov random field (MRF)-based method (ColluEagle) to detect collusive review spammers, as well as review spam campaigns, considering both network effects and time effects. First we identify co-review pairs, a review phenomenon that happens between two reviewers who review a common product in a similar way, and then model reviewers and their co-review pairs as a pairwise-MRF, and use loopy belief propagation to evaluate the suspiciousness of reviewers. We further design a high quality yet easy-to-compute node prior for ColluEagle, through which the review spammer groups can also be subsequently identified. Experiments show that ColluEagle can not only detect collusive spammers with high precision, significantly outperforming state-of-the-art baselines --- FraudEagle and SpEagle, but also identify highly suspicious review spammer campaigns.
\end{abstract}
\\


\section{Introduction}
\label{intro}
Online product reviews are increasingly influencing customers' purchase decisions, and thereby influencing product sales. To promote or demote product reputations, review spammers try to game the review websites by posting untruthful review content and/or rating stars. Ordinary customers have much difficulties in distinguishing fake reviews from genuine ones, as a result,  are vulnerable to review spamming. Nowadays, as the word-of-mouth marketing prevails, group spamming, i.e., a group of review spammers working together to promote or demote a set of target products, is becoming the new form of review spamming.

Over the years, researchers proposed various techniques to detect spam reviews, review spammers, or spammer groups. However, the problem is far from being solved because the underlying mechanism of review spamming is still unclear. Previous research focus on review content \cite{Jindal+Liu:08a,Ott_findingdeceptive,DBLP:conf/acl/LiCL13}, review behavior \cite{Limep2010,DBLP:conf/icwsm/MukherjeeV0G13}, and the relationships among reviewers, reviews and products \cite{DBLP:conf/icdm/WangXLY11,Mukherjee2012,kais-zwang}. These methods are shown to be effective in spotting certain kinds of spamming activities. Nonetheless, there is no one-size-fits-all solutions to detect all kinds of review spamming, due to the ever-changing spam strategies and the emerging review domains, etc. Therefore, it seems that the best way for detecting review spam is to incorporate as many approaches as possible.

Since it is hard to classify a review/reviewer as fake/real, ranking-based methods are often used to rank reviews/reviewers according to their suspiciousness in committing spam \cite{Limep2010,DBLP:conf/icdm/WangXLY11,conf/icwsm/AkogluCF13,kdd/RayanaA15}. Recently, Markov random field (MRF)-based methods, FraudEagle \cite{conf/icwsm/AkogluCF13} and SpEagle \cite{kdd/RayanaA15}, are shown to be superior to other ranking methods. FraudEagle is a light-weight detection method, which models the \textit{reviewer - product} bipartite network with signed edges as a MRF. To facilitate generality, FraudEagle does not use review content information and any prior knowledge of nodes. SpEagle builds on FraudEagle framework and extends it in two main directions: (1) Extending the graph representation to the \textit{reviewer - review - product} tripartite network, and (2) incorporating review/reviewer/product priors into the MRF network. It turns out that SpEagle is much more accurate than FraudEagle in detecting spam reviews and review spammers, although it takes considerable more efforts and time to compute reviewer priors, review priors, and product priors.

While FraudEagle and SpEagle both exploit review network effects to detect suspicious review/reviewers, they ignore time effects, which play a critical role in characterizing group spamming. Another drawback of these two methods is that they only rank individual review spammers, but are not able to detect review spammer groups. In this paper, considering both review network effects and time effects, we propose a novel MRF-based model that can detect both collusive review spammers and review spam campaigns in which they are involved. We identify that such collusive activities can be attributed to co-review pairs, a review phenomenon between two reviewers who write fake reviews towards a common product in a collusive way --- rating similar scores within a short period of time. We then model the co-reviewing behavior into a pairwise-MRF, with nodes representing reviewers and edges representing the co-review relationships. The contributions of our work are three-fold:
\begin{itemize}
\item We propose ColluEagle (resembling FraudEagle and SpEagle), a novel pairwise-MRF model which elegantly embeds the co-review phenomenon. We design a loopy belief propagation-based algorithm to infer the likely-hood of a reviewer being involved in group spamming.
\item We design for ColluEagle a computationally efficient reviewer prior, namely Neighbor tightness ($NT$), which can significantly boost the performance of ColluEagle. Meanwhile, the review spam campaigns are simultaneously identified and ranked through the spam scores of the individual campaign members evaluated by the MRF-model.
\item We conduct extensive experiments to evaluate the performance of ColluEagle. Experiments show that ColluEagle not only detects collusive reviewer spammers with significantly higher performance than FraudEagle and SpEagle, but also detects highly suspicious review spammer groups.
\end{itemize}

The remaining of this paper is organized as follows. Section~\ref{sec:rel} discusses the related work. In Section~\ref{sec:ColluEagle}, we first describe the pairwise-MRF model and its inference method, then we give the method to compute prior $NT$ and identify review spammer groups using the spamicity of individual group memebers. Section~\ref{sec:experiments} gives the experimental results. We conclude our work in Section~\ref{sec:concl}.

\section{Related work}
\label{sec:rel}
Jindal and Liu first proposed the fake review detection problem \cite{Jindal+Liu:08a}. The problem can be further categorized into fake review detection \cite{Jindal+Liu:08a,Ott_findingdeceptive,Xie2012,pkdd2015Ye}, fake reviewer detection \cite{Limep2010,DBLP:conf/icdm/WangXLY11,Xu:2013}, and review spammer group detection \cite{Mukherjee2012,pkdd2015Ye,comjnl-zwang,kais-zwang,apin-zwang}. There are also many survey papers \cite{Crawford2015,DBLP:journals/jikm/RastogiM17,DBLP:journals/widm/VivianiP17} summarizing the abundant works in this research field.

In recent years, detecting collusive review spammers or review spam campaigns is becoming a new trend. The first work aiming to detect review spammer groups is by Mukherjee et al. who use Frequent Item Set (FIM) mining to generate candidate reviewer groups, and design a PageRank-like algorithm (GSRank) to compute the spamicity of a reviewer group \cite{Mukherjee2012}. Xu et al. proposed a statistical model LCM (for Latent Collusion Model) and use EM algorithm to infer the collusion from the given FIM candidate colluder groups \cite{DBLP:conf/icdm/XuZ15}. These works rely on FIM to generate candidate suspicious reviewer groups or individual suspicious reviewers. FIM, however, is not capable of finding small reviewer groups (e.g., a group of review spammers only targeting one product) or loosely connected reviewer groups \cite{comjnl-zwang}.
\begin{figure}
  \scalebox{0.35}[0.35]{\includegraphics{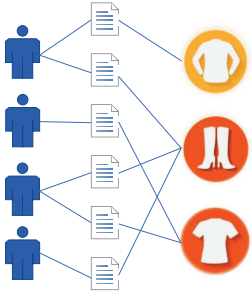}}
\caption{SpEagle graph representation}
\label{fig:speagle}
\end{figure}

The most closely related works to ours are FraudEagle and SpEagle, which exploit MRF to rank review spammers and fake reviews. FraudEagle models reviewers and products as MRF nodes, and the rating relations between reviewers and products as MRF edges. The edges are signed with``+'' or ``-'' representing a reviewer writing a positive or negative review for a product, which corresponds to different edge potentials in a pairwise-MRF. SpEagle extends from FraudEagle by also regarding the reviews as a type of nodes as shown in Fig.~\ref{fig:speagle}, with the \textit{reviewer - review} edge defined as ``write'', and the \textit{review - product} edge defined as ``belong to''. Therefore, the compatibility (edge) potentials are designed to reflect the fact that all the reviews written by a spammer (benign users) are all fake (genuine), and the majority of the reviews for targeted (non-targeted) products are fake (genuine). Although these assumptions make sense for fake reviews and review spammers in review data, both FraudEagle and SpEagle fail to reveal the collusion nature of group review spammers. So in our work, we introduce co-review pairs into our MRF model to capture collusive review spammers, which is notably different from FraudEagle and SpEagle. Moreover, FraudEagle and SpEagle themselves do not identify review spammer groups. To detect groups, further steps (e.g., clustering) are exploited upon the top-ranked reviewers, e.g., FraudEagle exploits the top ranked reviewers to obtain an induced subgraph of users and products, and the cross-associations clustering algorithm were used to detect bipartite cores. In contrast, our proposed ColluEagle detects individual review spammers and spammer groups in a holistic manner, and detects the review spammer groups over the whole review dataset.

There are also other group spamming detection works using MRF. For instance, Fei et al. first use kernel density estimation to detect review bursts in product reviews, and then use MRF and LBP to identify if a review burst is a normal burst (e.g., by TV commercial) or is under a spam attack \cite{fei2013exploiting}. They treat all the reviewers in a burst as a fully-connected graph (clique), and each reviewer can be in one of the three hidden states: non-spammer, mixed, and spammer. Although both use MRFs, ColluEagle does not rely on review bursts, and the underlying topological structures of the two MRF graphs are notably different. Li et al. also use MRF to model the relationships among twitter users, the URLs in tweets, and the burstiness of tweets \cite{DBLP:conf/icdm/LiMLKE14}. Therefore, a typed-MRF model is designed to detect Twitter campaign promoters. In their work, however, Twitter campaign promoters do not have to be twitter spammers. Moreover, their MRF model has three types of nodes and the edge semantics do not conform to those in review data environment.

\section{Methodology}
\label{sec:ColluEagle}
In this section, we first describe ColluEagle model and its LBP inference algorithm, then we discuss how to use node priors to boost the performance of ColluEagle, along with the method exploited to detect review spammer groups.
\subsection{Co-review pairs}

\begin{figure}
  \scalebox{0.45}[0.45]{\includegraphics{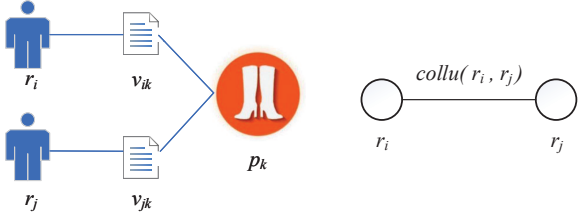}}
\caption{A co-review pair and the collusiveness between two reviewers}
\label{fig:coreview}
\end{figure}

Previous studies reveal that there are a large number of review spammers when a product suddenly receives a review burst \cite{fei2013exploiting,apin-zwang}. Such review bursts often involve group review spamming campaigns, for which there exist a group of review spammers working collaboratively to promote or demote a set of target products. To describe group spamming, we formulate this collective opinion spamming activity as a co-review behavior between two reviewers who write similar reviews towards one or more target products. Fig.~\ref{fig:coreview} illustrates a co-review pair for which reviewer $r_i$ and reviewer $r_j$ co-review a target product $p_k$, along with their corresponding reviews $v_{ik}$ and $v_{jk}$. The consistency of a co-review pair can be measured by the similarity between review $v_{ik}$ and $v_{jk}$, which usually consist of the review content and the meta-data such as review timestamps $t_{ik}$ and $t_{jk}$, and rating stars $\psi_{ik}$ and $\psi_{jk}$ for reviewer $r_i$ and $r_j$, respectively.

There shall be various methods to quantify the consistency of a co-review pair, here we give a measure of co-review similarity from a probabilistic point of view. As reported by many researchers \cite{Ott_findingdeceptive,DBLP:conf/icwsm/MukherjeeV0G13,kdd/RayanaA15}, review content (text) is unreliable in judging the truthfulness of a review, thus we only consider review meta-data (review date and rating stars).

\begin{myDef}
\label{def:co-review-sim}
\textit{Co-review similarity}: Given two reviewers $r_i$ and $r_j$ who co-review a product $p_k$ with their reviews $v_{ik}$ and $v_{jk}$, assume $\Delta t = t_{ik} - t_{jk}$, $\Delta \psi = \psi_{ik} - \psi_{jk}$ both follow a normal distribution, and $\Delta t \sim N(0,\sigma_1^2)$, $\Delta \psi \sim N(0,\sigma_2^2)$, we define the co-review similarity of $v_{ik}$ and $v_{jk}$ towards $p_k$ as:
\begin{equation}\label{eq:co-review-sim}
    \sigma(v_{ik},v_{jk},p_k) = 4\ \Phi(-\frac{|\Delta t|}{\sigma_1})\ \Phi(-\frac{|\Delta \psi|}{\sigma_2})
\end{equation}
where $\Phi(\cdot)$ denotes the cumulative distribution function (CDF) of the standard normal distribution.
\end{myDef}

The intuition of Definition~\ref{def:co-review-sim} is illustrated as follow. We assume random variable $\Delta t$ and $\Delta \psi$ follow normal distributions with parameters $N(0,\sigma_1^2)$ and $N(0,\sigma_2^2)$, respectively. Assuming $\Delta t$ and $\Delta \psi$ are independent, we use the joint probability of $\Delta t$ and $\Delta \psi$, $p(\Delta t,\Delta\psi)=p(\Delta t)p(\Delta\psi)$, to represent the similarity of the co-review pair. It is easy to see that the co-review similarity gets larger as their review time interval gets shorter and their rating stars get closer. Note that parameter $\sigma_1$ and $\sigma_2$ can be estimated from review data. Because each $\Phi(\cdot)$ in Eqn. \ref{eq:co-review-sim} is in $(0,0.5]$, $\sigma(v_{ik},v_{jk},p_k) \in (0,1]$.

Since two reviewers $r_i$ and $r_j$ might commonly review multiple target products, we then define the collusiveness between $r_i$ and $r_j$ by considering multiple co-review pairs, as illustrated in Fig.~\ref{fig:coreview}.
\begin{myDef}
\label{def:collu}
\textit{Collusiveness between two reviewers}: Given two reviewers $r_i$ and $r_j$, let
$P_i$ denote the product set reviewed by $r_i$, we define the collusiveness between $r_i$ and $r_j$ as:
\begin{equation}
\label{eq:collu}
collu(r_i,r_j) = \max_{p_k\in P_i\cap P_j}\sigma(v_{ik},v_{jk},p_k)
\end{equation}
\end{myDef}


\subsection{The ColluEagle model}
To compute the spamicity of each reviewer in a \textit{reviewer - review - product} tripartite network shown in Fig. \ref{fig:speagle}, we first construct a \textit{reviewer - reviewer} graph as defined in Definition~\ref{def:reviewer-graph}.
\begin{myDef}
\label{def:reviewer-graph}
\textit{Reviewer graph}: Let $R = \{r_1,r_2,\ldots,r_N\}$ be a set of $N$ reviewers. By computing the collusiveness of each \textit{reviewer - reviewer} pair using Equation \ref{eq:collu}, we can construct a reviewer graph $G = (V,E)$, where $V = R$, and $E=\{(r_i,r_j)|collu(r_i,r_j) \ge \delta, r_i, r_j \in R\}$, where $\delta$ is a threshold controlling the density of graph $G$.
\end{myDef}

In summary, reviewer graph $G$ not only captures the network effect of the \textit{reviewer - product} bipartite graph, but also captures the time effect that collusive reviewers often  co-review a common product in a narrow time window. This paves the way for us to identify suspicious reviewers from the huge amount of review data by using a Markov random field (MRF) model.

MRF is often used to model a set of random variables having a Markov property described by an undirected graph. In particular, a pairwise-MRF is a MRF satisfying the pairwise Markov property, i.e., a random variable is assumed to be dependent only on its neighbors and independent of all other variables. We then model $G$ as a pairwise-MRF by associating each node (reviewer) with a random variable $X_i, i = 1,\ldots,N$, where $X_i$ corresponds to $r_i \in R$. Let $\mathcal{L}=\{+1, -1\}$ denotes a set of labels, where $+1$ denote \textit{benign} and $-1$ denotes \text{spammer}, and $x_i \in \mathcal{L}$ denotes the label of node (or the value of random variable) $X_i$. As such, the review spammer detection problem can be formulated to a classification problem using a MRF based on reviewer graph $G$, i.e., inferring the class labels of all latent variables $X=\{X_1, X_2,\ldots,X_N\}$ given the observed reviewer graph $G$ and a set of spamicity priors for reviewers in $R$.

To build a complete pairwise-MRF model, we shall specify the node potentials of the class labels of node $X_i$, denoted as $\psi(x_i)$, and the edge potentials (or compatibility potentials) for all combinations of the labels of two adjacent nodes $X_i$ and $X_j$, denoted as $\psi(x_i,x_j)$. In this formulation, we define node potentials as the probabilities of a reviewer being a benign or spammer, i.e., $\psi(x_i) =  p(x_i)$. In MRF we can simply give $\psi(x_i)$ a fixed value, e.g., $(0.8,0.2)$ for labels $\{+1, -1\}$, or a prior derived from some features extracted from review data (see Section~\ref{sec:priors}).
Moreover, we define the edge potential between two adjacent nodes with labels $x_i$ and $x_j$ as
\begin{equation}\label{eq:edge potential}
    \psi(x_i,x_j) = e^{x_i x_j collu(r_i,r_j)}
\end{equation}
Equation \ref{eq:edge potential} implies that the probability a benign node having a benign neighbor, or a spammer node having a spammer neighbor is much larger than a benign node having a spammer neighbor, or a spammer node having a benign neighbor. Therefore, the joint probability of a configuration of the labels of all the nodes in a MRF can be written as:

\begin{equation}\label{eq:joint prob}
    p(X) = \frac{1}{Z}\prod_{X_i\in V} \psi(x_i) \prod_{(X_i,X_j) \in E} \psi(x_i,x_j)
\end{equation}
where $Z$ is the normalization constant. However, $p(X)$ is computationally intractable for large scale reviewer graphs, thus we resort to approximate inference methods to compute the suspiciousness of each reviewer in $G$.

\subsection{Model inference}
\label{sec:lbp}
To infer the node labels in reviewer graph $G$, we use loopy belief propagation (LBP), a widely adopted approximate inference algorithm for MRF. In LBP, a node iteratively sends messages to its neighbor nodes until all the messages become stationary. Let $m_{i\to j}(x_j)$ denote the message of label $x_j$ passed from node $X_i$ to node $X_j$, the message passing formula is:
\begin{equation}\label{eq:message-pass}
    m_{i\to j}(x_j)=\frac{1}{Z_1}\sum_{x_i \in \mathcal{L}}{\psi(x_i)\psi(x_i,x_j)\prod_{X_k \in N_i  \backslash X_j}m_{k\to i}(x_i)}
\end{equation}
where $N_i$ denotes the set of neighbor nodes of $X_i$, and $Z_1$ is the normalization constant. Equation~\ref{eq:message-pass} means that the message (benign or spammer) passed from node $i$ to node $j$ is proportional to the sum, over each $x_i \in \mathcal{L}$, of the product of the node potential of node $i$, the compatibility potential between $x_i$ and $x_j$, and all the messages passed from the neighbors of node $i$ except for node $j$ to node $i$. The above message update equation is called the Sum-product LBP.

In the initial phase of LBP, all the messages are set to 1. Then LBP iteratively updates each node using Equation~\ref{eq:message-pass}.
When LBP converges, the final belief that a node $X_i$ having label $x_i$ can be computed by:
\begin{equation}\label{eq:belief}
    b_i(x_i) = \frac{1}{Z_2}\psi(x_i)\prod_{X_k \in N_i}m_{k\to i}(x_i)
\end{equation}
where $Z_2$ is the normalization constant.

\subsection{Using priors}
\label{sec:priors}
From Equation \ref{eq:message-pass} and \ref{eq:belief} we can see that node priors in ColluEagle play a important role in ranking review spammers, which is also a main beauty of SpEagle. The difference in using priors is that SpEagle has three kinds of nodes --- reviewer nodes, review nodes, and product nodes, while ColluEagle only consists of reviewer nodes. In this subsection, we consider two reviewer priors --- prior $ALL$ and prior $NT$, for ColluEagle.

\subsubsection{Prior $ALL$}

\begin{table*}[!htbp]
\caption{Review-based behavioral features used in \cite{kdd/RayanaA15}}
\label{tbl:review-feature}
\begin{tabular}{lp{8cm}}
\hline\noalign{\smallskip}
Abbr. & description \\
\noalign{\smallskip}\hline\noalign{\smallskip}
Rank & Rank order among all the reviews of product\\
RD & Absolute rating deviation from product's average rating\\
EXT & Extremity of rating, 1 for \{5,4\}, 0 for \{1,2,3\}\\
DEV & Thresholded rating deviation of a review\\
ETF & Early time frame, spammers often review early\\
ISR & Is singleton review?\\
\noalign{\smallskip}\hline
\end{tabular}
\end{table*}
SpEagle takes into account three kinds of features --- reviewer-based features, product-based features, and review-based features, each of which is further categorized into textual features and behavioral features. As reported in \cite{kdd/RayanaA15}, the most effective features are the review-based features, and the product-based features have almost no effect in SpEagle. Table~\ref{tbl:review-feature} lists the review-based behavioral features.

To serve as the priors of each kind of nodes, in SpEagle, features of the same kind are preprocessed with the following two steps:
\begin{itemize}
\item Step 1: Normalize. To unify the features for the particular nodes into a comparable scale and interpretation, the empirical cumulative distribution function (CDF) is used whatever a high or low feature value is more suspicious. Specifically, for each feature $l$, $1 \le l \le F$, where $F$ is the total number of features for that kind of nodes, and the corresponding value of node $i$, denoted by $x_{li}$, we compute
    \begin{equation}
    f(x_{li})=\bigg\{
    \begin{array}{l}1-P(X_l\le x_{li}),\  if\ high\ is\ suspicious\\
        P(X_l\le x_{li}), \  otherwise
    \end{array} \nonumber
    \end{equation}
    where $X_l$ denotes a real-valued random variable associated with feature $l$ that follows probability distribution $P$. As a result, the lower the value of $f(x_{li})$, the more suspicious the node gets, whatever high or low the original feature values are.
\item Step 2: Combine. Given $F$ features of node $i$, the spam score of node $i$ is computed by combining all the normalized feature values of node $i$, i.e.,
    \begin{equation} \label{eq:combine}
    S_i = 1 - \sqrt{\frac{\sum_{l=1}^{F}f(x_{li})^2}{F}}
    \end{equation}
\end{itemize}

Based on the above two steps, we compute the prior for each review, namely prior $ALL$, by combining the six review-based features in Table~\ref{tbl:review-feature}. We do not use textual features because it is reported that textual features are computationally inefficient and perform poor in distinguishing review spam from truthful review \cite{kdd/RayanaA15,Xu:2013}. Our experimental study shows that this is a high quality prior for identifying both fake reviews and review spammers in Yelp datasets.

Since ColluEagle only has reviewer nodes, a straightforward scheme is to use the reviewer-based features to devise the reviewer node priors. However, as observed in SpEagle, reviewer-based features are not as effective as review-based features, at least in Yelp datasets. An alternative scheme is to transform review-based features to reviewer-based features by assuming that if a review is fake, then the author of the review is also a spammer. As such, for a reviewer $r_i$, we choose to set the prior of $r_i$ to the maximum prior of all the reviews written by reviewer $r_i$:
\begin{equation}
prior(r_i) = \max_{v_{ik} \in V_{r_i}} {prior(v_{ik})}
\end{equation}
where $V_{r_i}$ denotes the review set of $r_i$.
Our experiments show that this scheme can significantly improve detection precision in comparison to the scheme using the reviewer-based features proposed in SpEagle.

\subsubsection{Prior $NT$}

Since prior $ALL$ is expensive to compute, and domain knowledge-related, we resort to design an easy-to-compute and domain-independent node prior for ColluEagle by exploiting the topological structure of the review network that reflects the collusive behavior of review spammers. First we introduce the definition of a \textit{companion reviewer graph}.
\begin{myDef}
\label{def:companion-reviewer-graph}
\textit{Companion reviewer graph}: Let $R = \{r_1,r_2,\ldots,r_N\}$ be a set of $N$ reviewers. By computing the collusiveness of each \textit{reviewer - reviewer} pair using Equation \ref{eq:collu}, we can construct a companion reviewer graph $G^* = (V,E^*)$, where $V = R$, and $E^*=\{(r_i,r_j)|collu(r_i,r_j)\frac{|P_i \cap P_j|}{|P_i \cup P_j|} \ge \delta', r_i, r_j \in R\}$, where $\delta'$ is a threshold controlling the density of graph $G^*$.
\end{myDef}

The difference between a reviewer graph $G$ and its companion reviewer graph $G^*$ is that the edge weight between node $r_i$ and node $r_j$ is multiplied by the Jaccard similarity of reviewer $r_i$ and $r_j$ in $G^*$, thus the companion graph is much sparser than its original reviewer graph for $\delta = \delta'$. Note that $\delta'$ in a companion reviewer graph is not necessary to be equal to $\delta$ in its original reviewer graph. The Jaccard co-efficient restricts the two reviewers to review more products in common, and at the same time, review less products that are not in common, which indicates a collusive review behavior between the two reviewers. It is easy to see that this companion graph can be conveniently computed at the same time when computing the original reviewer graph. However, the companion reviewer graph does not work well for LBP due to its sparsity, but it facilitates LBP in designing a high quality reviewer prior.

Once the companion reviewer graph is computed, we can use SCAN \cite{DBLP:conf/kdd/XuYFS07} to mine all the dense clusters in it. These dense clusters, which characterize group spamming, are highly suspicious thus are supposed to be the \textit{candidate review spammer groups}. We first compute the \textit{Neighbor tightness} ($NT$) of a group $g$, and then we take this $NT$ value as the prior of the spamicity of each reviewer in that group.
\begin{equation}\label{eq:NT}
    NT(g) = \frac{\sum_{r_i,r_j \in g}{collu(r_i,r_j)\frac{|P_i \cap P_j|}{|P_i \cup P_j|}}}{{|g| \choose 2}}\frac{1}{1+e^{-(|g|-2)}}
\end{equation}
where $|g|$ denotes the number of reviewers in group $g$. The last term is a sigmoid function which penalizes small groups.

\subsection{Ranking review spammer groups}

Recall that when computing prior $NT$, we use SCAN to generate candidate collusive spammer groups. The spamicity of these groups can be measured by the prior $NT$ of that group. However, the spamicity of these groups can be further evaluated by the average spamicity of the group members computed by LBP using prior $NT$ (or other priors). We call this spamicity the posterior of a candidate group. The interesting point here is that the candidate groups facilitate LBP by providing $NT$ prior, and conversely, LBP facilitates candidate groups by providing the spam scores of the group members. Thus our method can detect both individual review spammers and review spammer groups, in a holistic fashion.

\subsection{The ColluEagle algorithm}

\begin{algorithm*}[!htbp]
  \renewcommand{\baselinestretch}{1.2}
  \footnotesize
  \caption{\textit{ColluEagle}}
  \label{alg:ColluEagle}
  \begin{algorithmic}[1]
  \INPUT{$\mathcal{D}$: Review metadata (reviewer id, product id, date, rating score);
  \\$\delta$: a threshold for generating the reviewer graph;
  \\$\delta'$: a threshold for generating the companion reviewer graph;
  \\$\sigma_1$, $\sigma_2$: parameters used in co-review similarity computation;
  \\$FS$: a foreign spam prior vector for reviewers, e.g., prior $ALL$}
  \OUTPUT{Ranked individual reviewers and review spammer groups}
  \DESCRIPTION
  \STATE{Construct reviewer graph $G=(V,E)$ using $\mathcal{D}$ and $\delta$, as well as the companion reviewer graph $G^*$ using $\delta'$;}
  \STATE{Use SCAN on $G^*$ to find candidate review spammer groups;}
  \STATE{Compute group spam indicator $NT$ for each candidate group $g$, i.e., spam(g) = $NT$;}
  \STATE{Set the spam prior of reviewer $r_i$, denoted as $S_i$, to spam($g$), $r_i \in g$;}
  \FOR{each node $X_i \in V$}
    \IF{Foreign prior is used}
        \STATE{$\psi(x_i)=(1-FS_i,FS_i)$;}
    \ELSE
        \STATE{$\psi(x_i)=(1-S_i,S_i)$;}
    \ENDIF
  \ENDFOR
  \FOR{each edge $e = (X_i, X_j) \in E$}
    \FOR{$x_i, x_j \in \mathcal{L}$}
      \STATE{$m_{i\to j}(x_j)=1$; $m_{j\to i}(x_i)=1$;}
    \ENDFOR
  \ENDFOR
  \FOR{each connected component $C=(V_c,E_c)$ in $G$}
      \REPEAT{}
        \FOR{each node $X_i \in V_c$}
          \FOR{each node $X_j \in N_i$}
            \FOR{$x_j \in \mathcal{L}$}
              \STATE{update $m_{i\to j}(x_j)$ using Equation~\ref{eq:message-pass};}
            \ENDFOR
          \ENDFOR
        \ENDFOR
      \UNTIL{all messages stop changing;}
      \FOR{each node $X_i \in V_c$ and $X_i$ has neighbors}
        \FOR{$x_i \in \mathcal{L}$}
          \STATE{Compute belief $b_i(x_i)$ using Equation~\ref{eq:belief};}
        \ENDFOR
      \ENDFOR
  \ENDFOR
  \RETURN{ranked reviewers based on belief $b_i(x_i=-1)$, and ranked review spammer groups based on the average belief of the group members;}
  \end{algorithmic}
  \renewcommand{\baselinestretch}{1.5}
\end{algorithm*}

In a nutshell, we illustrate ColluEagle in Algorithm \ref{alg:ColluEagle}. It takes as input the review metadata, the aforementioned parameters, and a foreign spam prior vector ($FS$) if any. As output, it ranks reviewers according to their spam scores, and review spammer groups based on their average spam scores of the group members.

Line 1 constructs the reviewer graph $G$ based on Equation \ref{eq:collu} with the specified graph pruning parameter $\delta$, and simultaneously compute the companion reviewer graph $G^*$ using $\delta'$. Line 2-4 first find candidate review spammer groups using SCAN from $G^*$, and compute the $NT$ prior for each group, and then these $NT$ priors are converted to the reviewer spam priors. Line 5-11 set the node priors in MRF as $NT$, or a foreign prior in $FS$, if any. Line 12-16 perform initialization of the messages for the labels on both directions of the edge in $G$ (set to 1). Line 17-32 traverse each connected component in $G$ to compute the spamicity of each reviewer in that component, which greatly facilitates the parallelization of LBP. For each connected components in $G$, Line 18-26 iteratively send messages from a node to one of its neighbors using Equation~\ref{eq:message-pass}, until all the messages stabilize, i.e., converged. Line 27-31 compute the beliefs of the two class labels for all the nodes using Equation~\ref{eq:belief}. Note that we ignore the isolated nodes who have no neighbors, because they do not participate in LBP. Finally, Line 33 returns the spam scores of reviewers, which correspond to the belief a reviewer being a spammer, and also the candidate review spammer groups are ranked based on the average belief of the group members.

Here we use a method similar to Algorithm $ConstructReviewerGraph$ proposed in~\cite{kais-zwang} to construct the reviewer graph, for which the time complexity is $O(e)$, where $e$ is the number of edges in the reviewer-product bipartite graph. From Algorithm \ref{alg:ColluEagle} we can see that the time complexity of ColluEagle is proportional to the number of edges in graph $G$, and the number of iterations $k$. That is, $T(n) = O(k|E|)$, where $n = |R|$, $E$ is the set of edges in $G$. Fortunately, both $k$ and $|E|$ can be lowered by setting a maximum number of iterations and/or specifying a larger $\delta$.

\section{Experimental study}
\label{sec:experiments}
\subsection{Datasets and compared baselines}
To evaluate the performance of ColluEagle, we use two labeled datasets, YelpNYC and YelpZip, crawled by Rayana et al. \cite{kdd/RayanaA15}, containing reviews for restaurants from Oct 2004 to Jan 2015. Table~\ref{tbl:datasets} illustrates the datasets we used. In each dataset, each review is labeled as either \textit{fake} or \textit{genuine}, according to the fake review filtering algorithm of Yelp.com~\cite{kdd/RayanaA15}. Since the datasets do not label reviewers, we label a reviewer as \textit{spammer} if and only if there exists at least one review written by that reviewer that is fake. Otherwise, the reviewer is labeled as \textit{benign}, which is also the same way used in \cite{kdd/RayanaA15}.
\begin{table}[!htbp]
\caption{Review dataset statistics}
\label{tbl:datasets}
\begin{tabular}{lrrrll}
\hline\noalign{\smallskip}
Dataset & \#Reviews & \#Reviewers & \#Products\\
\noalign{\smallskip}\hline\noalign{\smallskip}
YelpNYC & 359,052 & 160,225 & 923\\
YelpZip & 608,598 & 260,277 & 5,044\\
\noalign{\smallskip}\hline
\end{tabular}
\end{table}

\begin{figure}
\centering
  \subfigure[YelpNYC]{
  \includegraphics[width=3in]{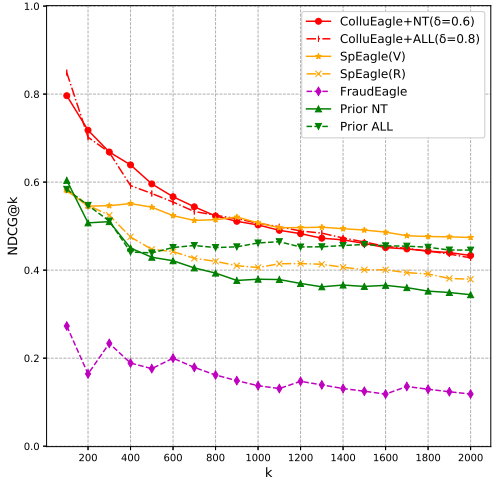}
  }
  \subfigure[YelpZip]{
  \includegraphics[width=3in]{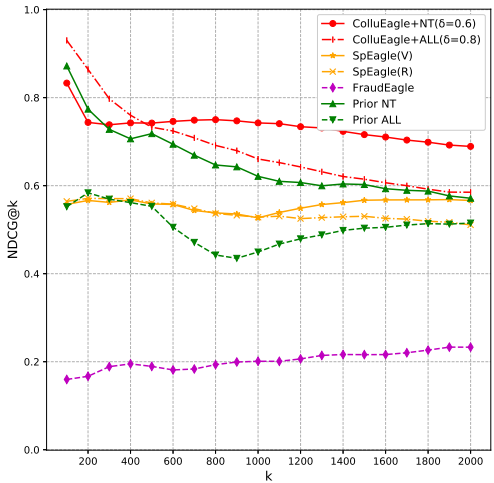}
  }
  \caption{Reviewer ranking comparison for ColluEagle, SpEagle, FraudEagle, and priors. ColluEagle+NT($\delta=0.6$) denotes ColluEagle using prior $NT$ and $\delta=0.6$, ColluEagle+ALL($\delta=0.8$) denotes ColluEagle using prior $ALL$ and $\delta=0.8$. SpEagle(R) denotes ranking reviewers according to reviewer nodes in SpEagle, and SpEagle(V) ranking reviewers according to review nodes in SpEagle. Here we set $\delta'=0.5$.}
  \label{fig:userPrior} 
\end{figure}

The most related works to ColluEagle are FraudEagle and SpEagle, which also rank reviewers and reviews by spam scores. Our proposed ColluEagle only contains reviewer nodes, thus it is best to use it to rank reviewers. To evaluate the performance of review spammer groups ranked by ColluEagle, we compare ColluEagle with our recent work for reviewer spammer group ranking, namely GSLDA \cite{apin-zwang}, which was shown to be superior to other group spammer detection methods.

\subsection{Performance for ranking individual reviewers}
\label{subsec:rankingreviewer}
First, we evaluate the performance of ColluEagle, FraudEagle and SpEagle on dataset YelpNYC and YelpZip for ranking individual reviewers. For ColluEagle we set $\sigma_1 = 90$, $\sigma_2 = 3$ for both datasets, and use prior $ALL$ or prior $NT$ as the node priors, respectively. We vary $\delta$ and $\delta'$ to generate various ranking lists of ColluEagle. For FraudEagle, we set the priors of reviewer nodes and product nodes to (0.5, 0.5) as the way done in FraudEagle. For SpEagle, we use the SpLite version which is computationally efficient and yields a similar performance to SpEagle. That is, we use prior $ALL$ as the review node priors, and set the priors of reviewer nodes and product nodes as (0.5, 0.5).

Since SpEagle has both reviewer nodes and review nodes, we can rank reviewers either from the reviewer node beliefs (denoted as SpEagle(R)) or from the review node beliefs (denoted as SpEagle(V)), assuming that a reviewer is a spammer if and only if he/she has written at least one fake review.

Since ColluEagle, SpEagle, and FraudEagle are all ranking-based approaches, we use NDCG@k to evaluate the top-ranked reviewers. NDCG (normalized discounted cumulative gain) is a metric to inspect the quality (precision) at the top of a ranking list, which is widely used in outlier/spam/fraud detection \cite{kdd/RayanaA15}. At this point, we consider two versions of ColluEagle: ColluEagle+ALL ($\delta = 0.8$), and ColluEalge+NT ($\delta = 0.6$). We set $\delta'=0.5$ for both YelpNYC and YelpZip to generate the companion reviewer graph $G^*$.

For YelpNYC and YelpZip dataset, Fig.~\ref{fig:userPrior} shows the NDCG@k comparison of the ranking results of ColluEagle, SpEagle, FraudEagle, prior $ALL$ and prior $NT$, each considering the top-ranked 2000 reviewers. We can see that, in general, ColluEagle significantly outperforms FraudEagle and SpEagle, especially for YelpZip dataset, indicating the distinct advantage of co-review pair-based method in detecting review spammers. As expected, FraudEagle yields the worst performance, for it does not leverage any prior knowledge of any node. In general, SpEagle(V) performs better than SpEagle(R), which reveals that review nodes play a more important role than reviewer nodes, probably because only the review priors are used. We also rank reviewers according to prior $ALL$ or prior $NT$. From Fig.~\ref{fig:userPrior} we can see that the two priors perform quite different for YelpNYC and YelpZip dataset, and sometimes even beat SpEagle, but none of them can exceed ColluEagle for both datasets.

\begin{figure}
\centering
  \subfigure[YelpNYC]{
  \includegraphics[width=2.2in]{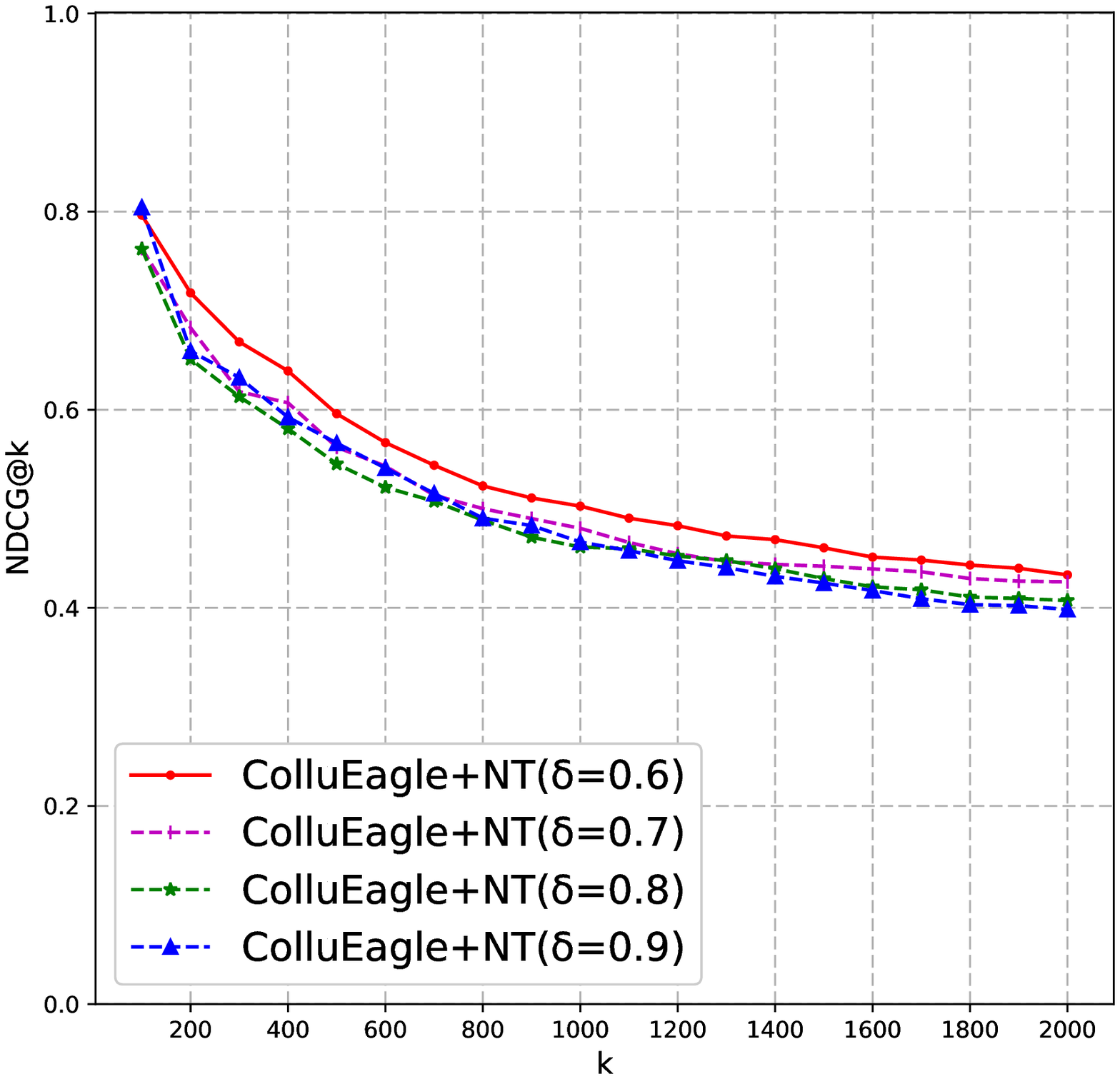}
  }
  \subfigure[YelpZip]{
  \includegraphics[width=2.2in]{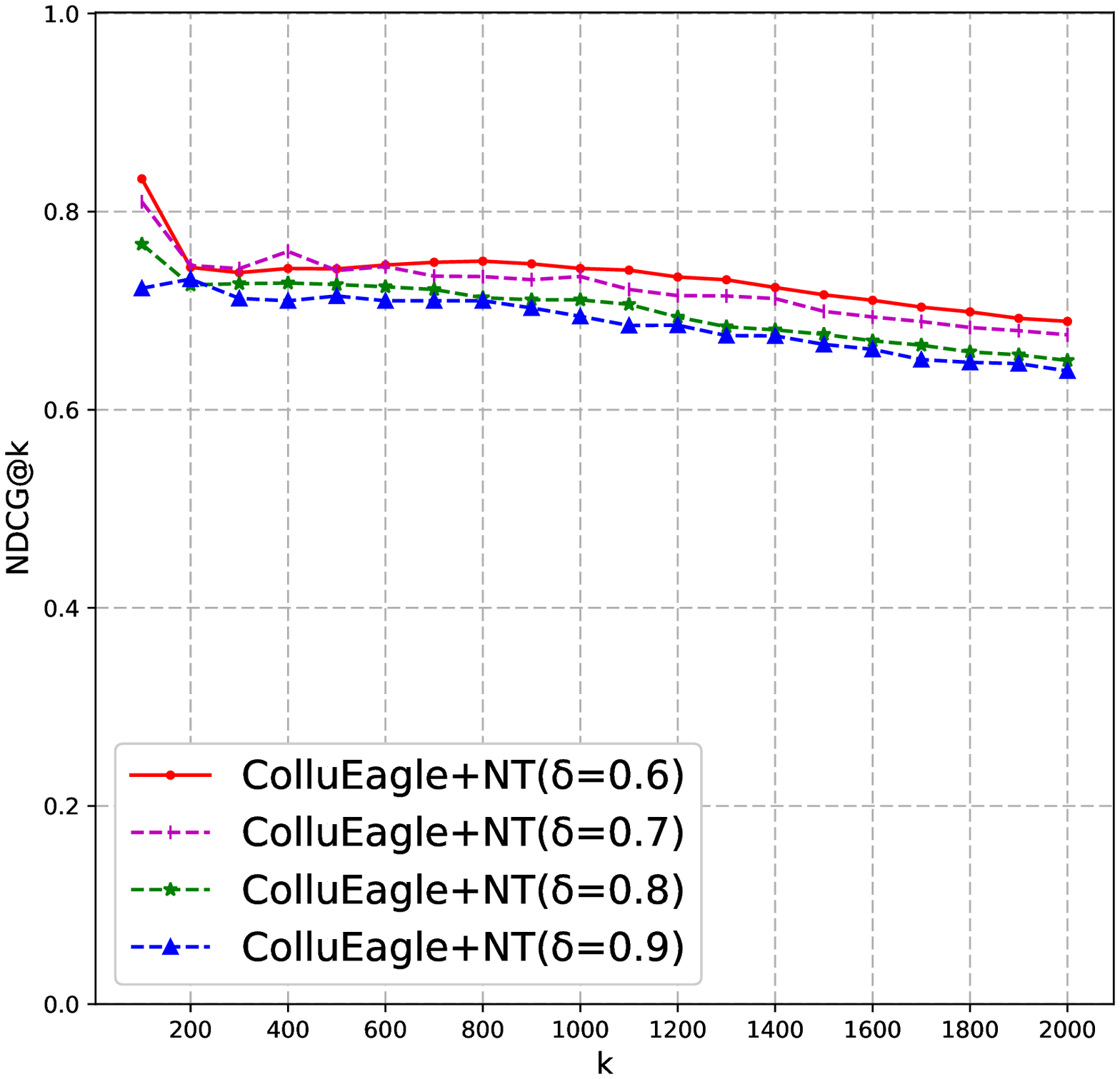}
  }
  \caption{The impact of parameter $\delta$ in ColluEagle ($\delta'=0.5$)}
  \label{fig:impact} 
\end{figure}

\begin{figure}
\centering
  \subfigure[YelpNYC]{
  \includegraphics[width=2.2in]{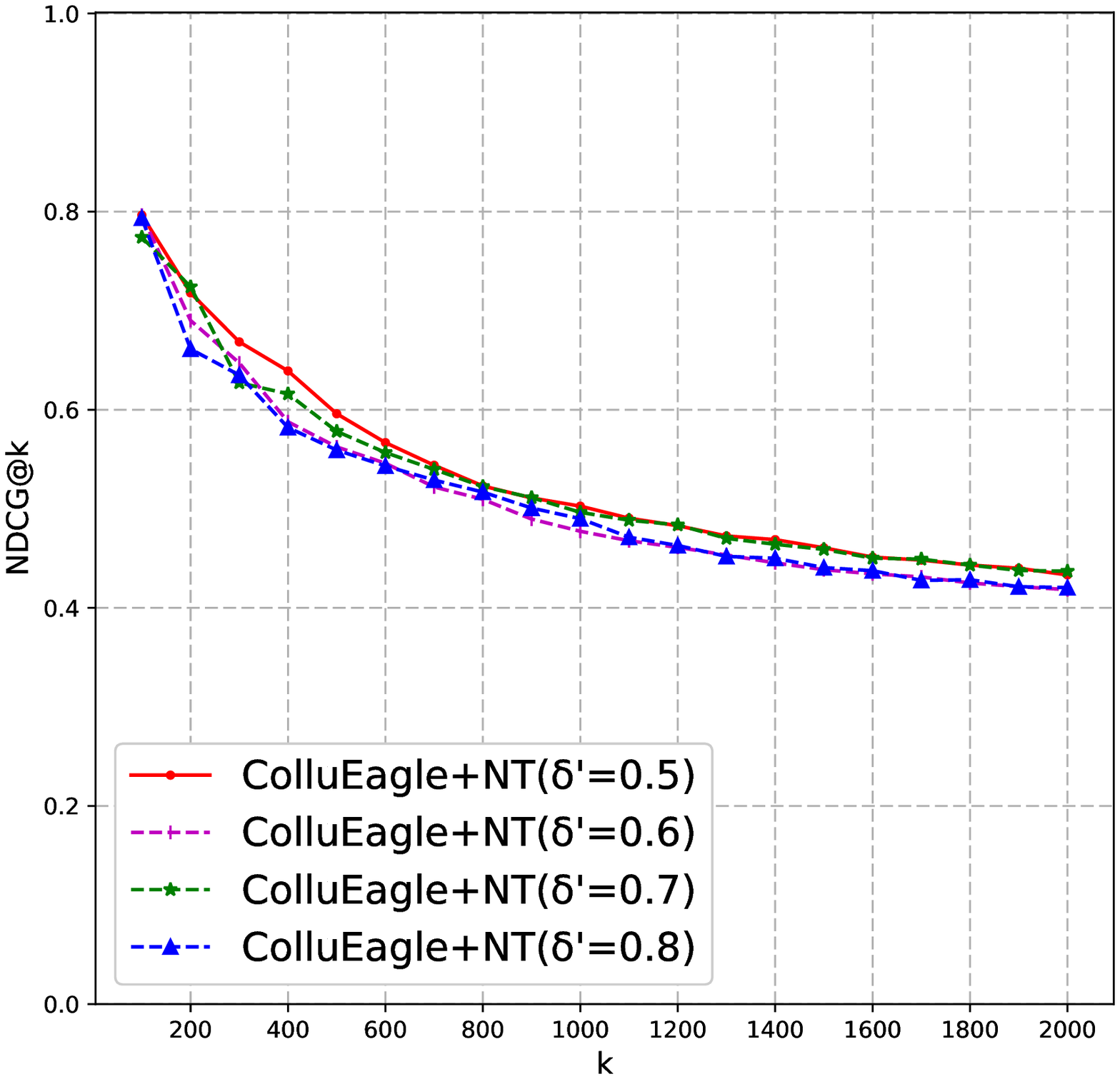}
  }
  \subfigure[YelpZip]{
  \includegraphics[width=2.2in]{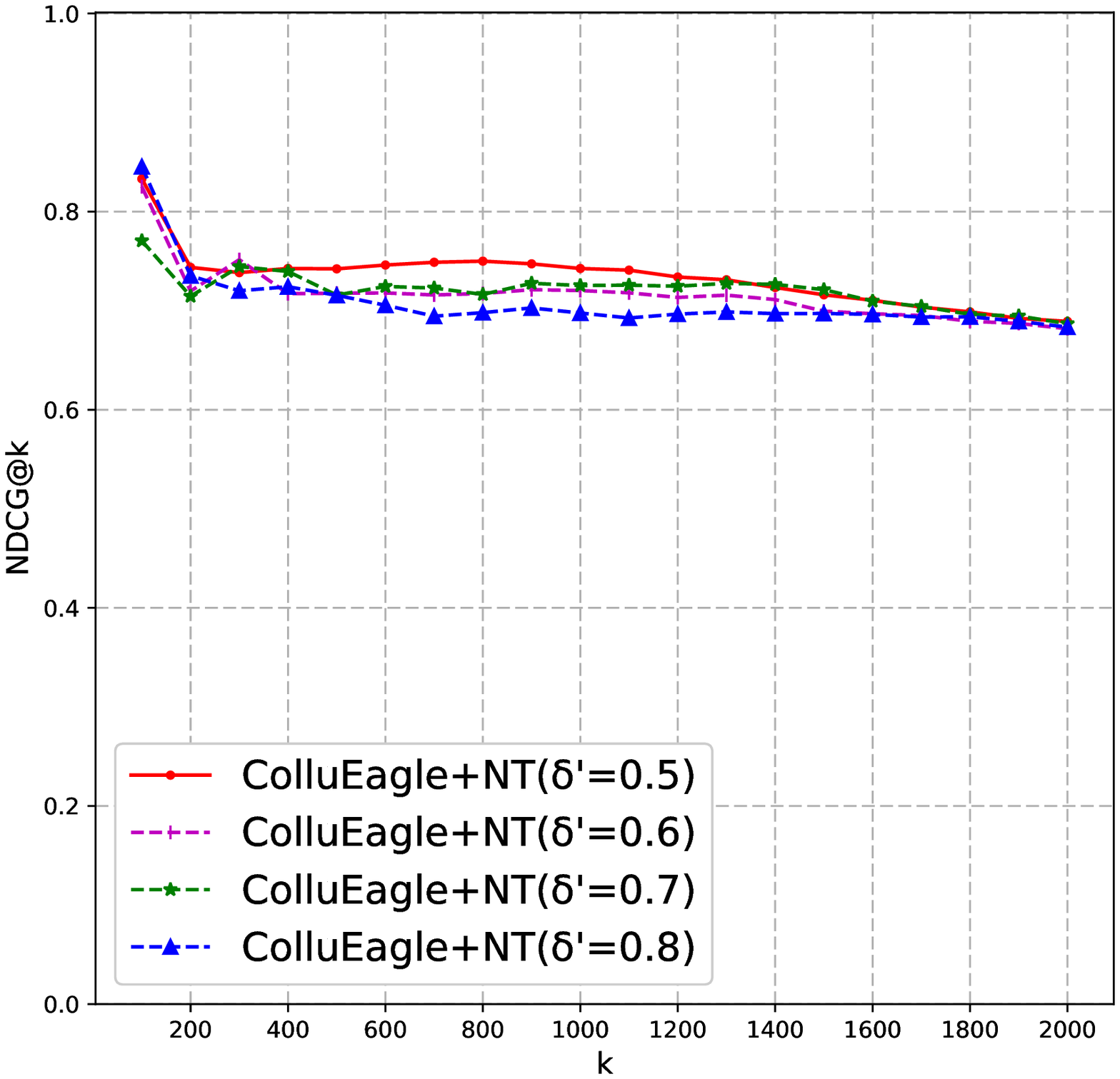}
  }
  \caption{The impact of parameter $\delta'$ in ColluEagle ($\delta=0.6$)}
  \label{fig:impact2} 
\end{figure}

As we can see, parameter $\delta$ has much impact on the performance of ColluEagle by influencing the underlying graph topology in MRF. Therefore we run ColluEagle+NT with different $\delta$, and keep $\delta'=0.5$, as shown in Fig.~\ref{fig:impact}. We can see that the performance for different $\delta$s are quite similar, indicating that ColluEagle is robust to $\delta$. In general, smaller $\delta$ yields better precision. This is because smaller $\delta$ produces more edges in the reviewer graph, which play a important role in LBP. Yet a small $\delta$ yields a large amount of edges, which slows down LBP.

Also we can see that $\delta'$ can impact the performance of ColluEagle by providing different prior $NT$s. To study the impact of $\delta'$, we run ColluEagle+NT ($\delta=0.6$) for different $\delta'$, as shown in Fig.~\ref{fig:impact2}. Again we can see that the performance does not change much as $\delta'$ varies from $0.5$ to $0.8$, and a smaller $\delta'$ would be preferred.  Experiments show that parameter $\sigma_1$ and $\sigma_2$ only slightly impact the detection precision, so we omit these results.

\begin{figure}
\centering
  \subfigure[YelpNYC]{
  \includegraphics[width=3in]{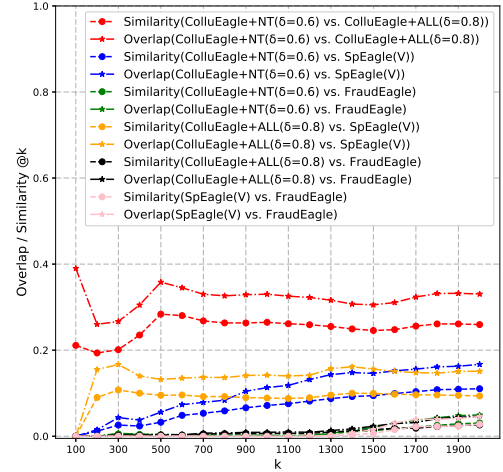}
  }
  \subfigure[YelpZip]{
  \includegraphics[width=3in]{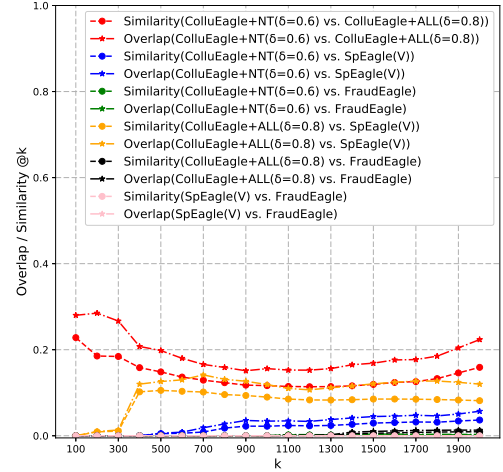}
  }
  \caption{Ranking list overlapping and similarity degree comparison among ColluEagle+NT($\delta=0.6$), ColluEagle+ALL($\delta=0.8$), SpEagle(V), and FraudEagle}
  \label{fig:Overlap} 
\end{figure}

As review spam varies from one to another in characteristics and strategies, different approaches might good at detecting different kinds of review spam. Therefore, we compared the ranking lists of ColluEagle, SpEagle, and FraudEagle. To measure the similarity between two ranking lists, we introduce two metrics called overlapping degree and similarity degree.

\begin{myDef}
\label{def:overlap}
\textit{Overlapping degree@k}: given two ranking lists $A$ and $B$, each consisting of $k$ reviewers, the overlapping degree of $A$ and $B$ is defined as:
\begin{equation}\label{eq:overlap}
    O(A,B) = \frac{|A\cap B|}{k}
\end{equation}
\end{myDef}

\begin{myDef}
\label{def:similarity}
\textit{Similarity degree@k}: given two ranking lists $A$ and $B$, each consisting of $k$ reviewers, let $a \in A$, $loc_A(a)$ denotes the position of $a$ in $A$ (start from 1), we define the distance from $a$ to $B$ as:
\begin{equation}\label{eq:dist}
    dist(a,B) = \bigg\{
    \begin{array}{l}|loc_A(a)-loc_B(a)|, if\ a\in B\\
        k,\ otherwise
    \end{array} \nonumber
\end{equation}
Then we define the similarity degree of $A$ and $B$ as:
\begin{equation}
S(A,B) = 1 - \frac{\sum_{a\in A}dist(a,B)}{k^2}
\end{equation}
\end{myDef}

\begin{table}[!htbp]
\caption{Top 2000 Ranking lists comparison of ColluEagle+NT($\delta'=0.5$) for different $\delta$ using YelpZip. Overlapping degree in upper-triangle, and Similarity degree in lower-triangle.}
\label{tbl:delta impact}
\begin{tabular}{l|rrrr}
\hline\noalign{\smallskip}
$\delta$ & 0.6 & 0.7 & 0.8 & 0.9\\
\noalign{\smallskip}\hline\noalign{\smallskip}
0.6 &  & 0.8375 & 0.7460 & 0.7245\\
0.7 & 0.7826 & & 0.8660 &0.7595\\
0.8 & 0.6720 & 0.8046 & & 0.8400\\
0.9 & 0.6440 & 0.6772 & 0.7694 &\\
\noalign{\smallskip}\hline
\end{tabular}
\end{table}

\begin{table}[!htbp]
\caption{Top 2000 ranking lists comparison of ColluEagle+NT($\delta =0.6$) for different $\delta'$ using YelpZip. Overlapping degree in upper-triangle, and Similarity degree in lower-triangle.}
\label{tbl:delta prime impact}
\begin{tabular}{l|rrrr}
\hline\noalign{\smallskip}
$\delta'$ & 0.5 & 0.6 & 0.7 & 0.8\\
\noalign{\smallskip}\hline\noalign{\smallskip}
0.5 &  & 0.7790 & 0.6640 & 0.5910\\
0.6 & 0.7300 & & 0.7695 & 0.6695\\
0.7 & 0.5941 & 0.7222 & & 0.7340\\
0.8 & 0.4883 & 0.5788 & 0.6667 & \\
\noalign{\smallskip}\hline
\end{tabular}
\end{table}

Clearly, $O(A,B)=O(B,A)$, $S(A,B)=S(B,A)$. We plot the overlapping and similarity degree@2000 among ColluEagle+NT($\delta=0.6$), CollEagle+ALL($\delta=0.8$), SpEagle(V) and FraudEagle, as shown in Fig.~\ref{fig:Overlap}. We can see that the overlapping and similarity degree between any two different methods are all very low. Even for the same algorithm, ColluEagle, with different priors ($ALL$ and $NT$), the similarity degree is below 0.3 for YelpNYC dataset,and 0.2 for YelpZip dataset. We also compared the overlapping and similarity degree between two ranking lists (top 2000 reviewers) of ColluEagle+NT with different $\delta$ (fixing $\delta'=0.5$), as shown in Table~\ref{tbl:delta impact}, and two ranking lists of ColluEagle+NT with different $\delta'$ (fixing $\delta=0.6$), as shown in Table~\ref{tbl:delta prime impact}, for YelpZip dataset. We can see that the overlapping and similarity degree decrease as the difference of $\delta$ or $\delta'$ increases. The overlapping degree ranges from 59\% to 87\%. This implies that these algorithms are good at detecting different kinds of review spammers, thus we can use different algorithms (ColluEagle, SpEagle), different priors (e.g., prior $ALL$ or $NT$), or even different $\delta$ and $\delta'$ to detect more review spammers all with a high precision, which can greatly improve the detection recall.

\begin{figure}
\centering
  \subfigure[YelpNYC]{
  \includegraphics[width=3in]{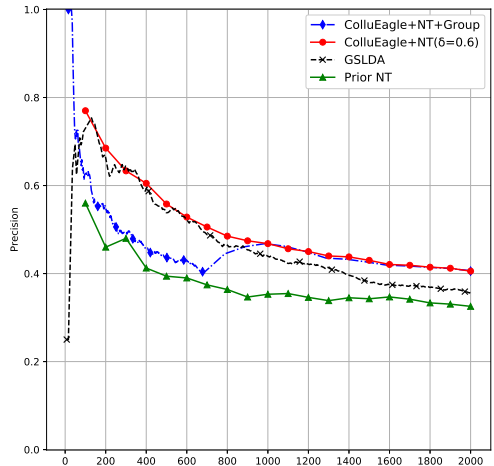}
  }
  \subfigure[YelpZip]{
  \includegraphics[width=3in]{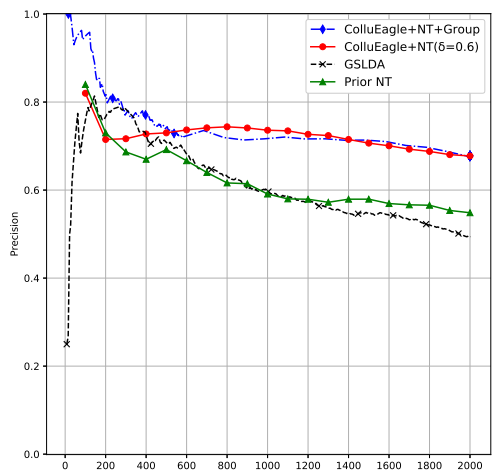}
  }
  \caption{Group ranking comparison on two datasets. For ColluEagle+NT+Group and GSLDA, a marker \textit{diamond} or 'x' is plotted every 40 groups. For ColluEagle+NT($\delta=0.6$) or prior $NT$, a marker is plotted every 100 reviewers.}
  \label{fig:grouprank} 
\end{figure}

\subsection{Performance for ranking review spammer groups}

Now we consider the performance of ranking review spammer groups detected by ColluEagle. Here we compare with GSLDA~\cite{apin-zwang}, a LDA-based method to detect review spammer groups. We use ColluEagle+NT($\delta=0.6$) to rank candidate review spammer groups by the average belief of all the members of that group. We tune the parameters in GSLDA so that the best performance is achieved. We fetch 2000 suspicious reviewers from the top ranked groups generated by ColluEagle using prior $NT$, $\delta=0.6$, $\delta'=0.5$ (denoted as ColluEagle+NT+Group) and GSLDA.

Fig.~\ref{fig:grouprank}(a) and (b) shows the detect precision of ColluEagle+NT+Group and GSLDA on YelpNYC and YelpZip, respectively. To measure the promotion of LBP in ColluEagle, we also plot the precision of the prior $NT$ and ColluEagle+NT($\delta=0.6$) for the top ranked 2000 reviewers.
We can see that, for YelpNYC, GSLDA outperforms ColluEagle+NT+Group for approximately the top 900 reviewers, but after that, ColluEagle+NT+Group outperforms GSLDA. ColluEagle+NT($\delta=0.6$) always outperforms GSLDA. For YelpZip, we can see that both ColluEagle+NT+Group and ColluEagle+NT($\delta=0.6$) outperform GSLDA by a large margin. To our surprise, ColluEagle+NT+Group achieves significantly higher precision than ColluEagle+NT($\delta=0.6$) for top 500 reviewers. We also can see that prior $NT$ is the overall looser, and significantly lower than ColluEagle+NT+Group, indicating that LBP can significantly improve the ranking quality of candidate review group spammers.

Previous reviewer spammer group detection methods, e.g. GSRank, GSLDA, etc., require computing several group spamming indicators or group member spamming indicators, which is computationally expensive and domain-related. In comparison, ColluEagle does not require computing any extra spam indicators, and can detect review spammers and groups in a holistic manner with a higher precision.

\section{Conclusion}
Detecting fake reviews or fake review spammers is a challenging problem and has attracted enormous research interest in recent years.
In this paper, we propose ColluEagle, a Markov random field-based detection method, to detect collusive review spammers, along with review spammer groups, by exploiting the co-review behavior. The method is completely unsupervised, time-efficient, and can detect different kinds of spamming behaviors under the guidance of prior knowledge extracted from review data. Experimental study shows that ColluEagle can dramatically improve the detection precision, outperforming state-of-the-art baselines (FraudEagle, SpEagle, and GSLDA) by a large margin, e.g., the NDCG@1000 metric for ColluEagle is improved about 20\% compared to SpEagle on YelpZip dataset. Future work include seeking new robust node priors, new effective co-review similarity metrics. A promising direction is to detect fake reviews in an online fashion, i.e., detecting emerging reputation fraud campaigns using MRF-based models. For reproducibility, we share our codes on Github\footnote{https://github.com/zhuowangsylu/ColluEagle}.
\label{sec:concl}


\bibliographystyle{spmpsci}      
\bibliography{spamgroup}   

\end{document}